\documentclass[aps,pre,twocolumn,groupedaddress]{revtex4}

\usepackage{algorithm}
\usepackage{algorithmic}

\ifx\pdfoutput\undefined
\usepackage{graphicx}
\else
\usepackage[pdftex]{graphicx}
\fi

\begin{document}

\title{Nature-Inspired Interconnects for Self-Assembled Large-Scale
  Network-on-Chip Designs}
\author{Christof Teuscher}
\email[]{christof@teuscher.ch}
\homepage[]{http://www.teuscher.ch/christof}
\affiliation{Los Alamos National Laboratory, CCS-3, MS-B256, Los Alamos, NM
87545}

\begin{abstract}
Future nano-scale electronics built up from an Avogadro number of
components needs efficient, highly scalable, and robust means of
communication in order to be competitive with traditional silicon
approaches. In recent years, the Networks-on-Chip (NoC) paradigm
emerged as a promising solution to interconnect challenges in
silicon-based electronics. Current NoC architectures are either highly
regular or fully customized, both of which represent implausible
assumptions for emerging bottom-up self-assembled molecular
electronics that are generally assumed to have a high degree of
irregularity and imperfection. Here, we pragmatically and
experimentally investigate important design trade-offs and properties
of an irregular, abstract, yet physically plausible 3D small-world
interconnect fabric that is inspired by modern network-on-chip
paradigms. We vary the framework's key parameters, such as the
connectivity, the number of switch nodes, the distribution of long-
versus short-range connections, and measure the network's relevant
communication characteristics. We further explore the robustness against
link failures and the ability and efficiency to solve a simple toy
problem, the synchronization task. The results confirm that (1)
computation in irregular assemblies is a promising and disruptive
computing paradigm for self-assembled nano-scale electronics and (2)
that 3D small-world interconnect fabrics with a power-law
decaying distribution of shortcut lengths are physically plausible and
have major advantages over local 2D and 3D regular topologies.
\end{abstract}


\maketitle

\section{Introduction}
\label{sec:intro}
It is generally expected that without disruptive new technologies, the
ever-increasing computing performance (commonly known as Moore's law
\cite{moore65}) and the storage capacity achieved with existing
technologies will eventually reach a plateau \cite{kish02}. However,
there is a lack of consensus on what type of technology and computing
architecture holds most promises to keep up the current pace of
progress. Among the most contemplated future and emerging technologies
are quantum computers, molecular electronics, nano-electronics,
optical computers, and quantum-dot cellular automata (QCA). In this
paper, we will primarily focus on self-assembled nano-scale
electronics based on nanowires or nanotubes because these fabrication
technologies have become quite mature on the physical and device
level. It is, however, still unclear how to develop higher-level
computational architectures in a reliable way, although a number of
promising approaches have been explored in detail (e.g.,
\cite{Heath98,patwardhan06,dehon03,tour02}) As Chen et
al. \cite{chen03} state, ``[i]n order to realize functional
nano-electronic circuits, researchers need to solve three problems:
invent a nanoscale device that switches an electric current on and
off; build a nanoscale circuit that controllably links very large
numbers of these devices with each other and with external systems in
order to perform memory and/or logic functions; and design an
architecture that allows the circuits to communicate with other
systems and operate independently on their lower-level details.''

Building a scalable computing architecture on top of a potentially
very unreliable physical substrate, such as for example molecular
electronics, is a challenging task, which is guided by a number of
major trade-offs in the design space \cite{stan03}, such as the number
and the characteristics of the resources available, the required
performance, the energy consumption, and the reliability. The lack of
systematic understanding of these issues and of clear design
methodologies makes the process still more of an art than of a
scientific endeavor and the appearance of novel and non-standard
physical computing devices (e.g., \cite{tour02}) generally only
aggravates these difficulties.

In recent years, the importance of interconnects on electronic chips
has outrun the importance of transistors as a dominant factor of
performance \cite{meindl03,ho01,davis01}. The reasons are twofold: (1)
the transistor switching speed for traditional silicon is much faster
than the average wire delays and (2) the required chip area for
interconnects has dramatically increased. The ITRS roadmap
\cite{itrs2003} lists a number of critical challenges for
interconnects and states that ``[i]t is now widely conceded that
technology alone cannot solve the on-chip global interconnect problem
with current design methodologies.'' The major challenges are related
to delays of non-scalable global interconnects and reliability in
general, which leads to the observation that simple scaling will no
longer satisfy performance requirements as feature sizes continue to
shrink \cite{ho01}.

In this paper, we experimentally and pragmatically investigate a
certain class of irregular and physically plausible 3D interconnect
fabrics, which are likely to be easily and cheaply implementable by
self-assembling either nanowires or nanotubes. We will vary the
framework's key parameters, such as the connectivity, the number of
switch nodes, the distribution of long- versus short-range
connections, and measure the network's relevant communication
characteristics and the robustness against failures. Further, we will
compare its performance with regular and nearest-neighbor connected 2D
and 3D cellular-automata-like interconnect fabrics. We will also
evaluate and compare the performance of a simple task that is
frequently used in the cellular automata community, the
synchronization task.

The motivation for investigating alternative and more
biologically-inspired interconnects, that can be self-assembled easily
and cheaply, can be summarized by the following observations: 

\begin{itemize}
  \item long-range and global connections are costly (in terms of wire
    delay and of the chip area used) and limit system performance
    \cite{ho01};
  \item it is unclear whether a precisely regular and homogeneous
    arrangement of components is needed and possible on a
    multi-billion-component or even Avogadro-scale assembly of
    nano-scale components \cite{tour02}
  \item ``[s]elf-assembly makes it relatively easy to form a random
    array of wires with randomly attached switches'' \cite{zhirnov01};
    and
  \item building a perfect system is very hard and expensive
\end{itemize}

By using an abstract, yet physically plausible and
fabrication-friendly nanoscale computing framework, we show that
self-assembled interconnect fabrics with small-world \cite{watts98}
properties have major advantages in terms of performance and
robustness over purely regular and nearest-neighbor connected fabrics,
such as cellular-automata-like topologies (sometimes called {\em NEWS}
communication, standing for north, east, west, south). While there is
ample evidence of the superior communication characteristics of
small-world and power-law over locally connected topologies (see also
Section \ref{sec:noc}), most abstract models are not physically
plausible and are thus of limited significance for real-world
implementations. For example, it is very unrealistic to assume a
uniform rewiring probability (as in the original Watts-Strogatz model
\cite{watts98}) over all possible nodes. Spatial aspects of
small-world topologies and wiring-cost perspectives have only recently
gained more attention
\cite{petermann05,petermann06,kozma05,kleinberg00,jespersen00}. We
call such interconnect topologies {\em nature-} or {\em bio-inspired}
because they are physically plausible and similar network topologies
are widespread in natural systems.

The goal of this paper is to experimentally investigate important
design trade-offs of self-assembled interconnect fabrics for emerging
nano-scale electronics. The main contribution consists in a pragmatic
comparison of regular (both 2D and 3D) versus irregular small-world
topologies of physically plausible self-assembled network-on-chip
(NoC) interconnect fabrics that are inspired by natural networks. We
believe that the results will help to make important design decisions
for future bottom-up self-assembled computing architectures. The
question of how much interconnect we need and how one can efficiently
build---or rather self-assemble---it, is a very important question,
not only for chip design and future molecular electronics in
particular, but also for any massively parallel computing architecture
in general.

The remainder of the paper is as following: Section \ref{sec:noc}
provides a brief introduction to complex networks and modern
network-on-chip (NoC) paradigms. Section \ref{sec:framework} describes
the framework that we use, such as the topologies used, the wire and
node model, and how physically plausible the approach is. Section
\ref{sec:experiments} reports on five simple experiments which
illustrate the main findings, while Section \ref{sec:conclusion}
finally concludes the paper.

\section{Networks and Networks-on-Chip}
\label{sec:noc}

\subsection{Complex Networks and Wiring Costs}
\label{sec:complexnets}
Most real networks, such as brain networks \cite{sporns04,egueluz05},
electronic circuits \cite{cancho01}, the Internet, and social networks
share the so-called {\em small-world} (SW) property
\cite{watts98}. Compared to purely locally and regularly
interconnected networks (such as for example the cellular automata
interconnect), small-world networks have a very short average distance
between any pair of nodes, which makes them particularly interesting
for efficient communication.

The classical Watts-Strogatz small-world network \cite{watts98} is
built from a regular lattice with only nearest neighbor
connections. Every link is then rewired with a {\em rewiring
  probability} $p$ to a randomly chosen node. Thus, by varying $p$,
one can obtain a fully regular ($p=0$) and a fully random ($p=1$)
network topology. The rewiring procedure establishes ``shortcuts'' in
the network, which significantly lower the average distance (i.e., the
number of edges to traverse) between any pair of nodes. In the
original model, the length distribution of the shortcuts is uniform
since a node is chosen randomly. If the rewiring of the connections is
done proportional to a power law, $l^{-\alpha}$, where $l$ is the wire
length, then we obtain a {\em small-world power-law network}. The
exponent $\alpha$ affects the network's communication characteristics
\cite{kozma05} and navigability \cite{kleinberg00}, which is better
than in the uniformly generated small-world network. One can think of
other distance-proportional distributions for the rewiring, such as
for example a Gaussian distribution, which has been found between
certain layers of the rat's neocortical pyramidal neurons
\cite{hellwig00}. Studying the connection probabilities and the
average number of connections in biological systems, especially in
neural systems, can give us important insights on how nearly optimal
systems evolved in Nature under limited resources and various other
physical constraints.

In a real network, it is fair to assume that local connections have a
lower cost (in terms of the associated wire-delay and the area
required) than long-distance connections.  Physically realizing
small-world networks with uniformly distributed long-distance
connections is thus not realistic and distance, i.e., the wiring cost,
needs to be taken into account, a perspective that recently gained
increasing attention \cite{petermann05,petermann06}. On the other
hand, a network's topology also directly affects how efficient
problems can be solved. For example, it has been shown that both
small-world \cite{tomassini05} as well as random Erd\"os-R\'enyi
topologies \cite{mesot05} offer better performance than regular
lattices and are easier to evolve to solve the global synchronization
and density classification task, two toy problems commonly used in the
cellular automata community.

In summary: there is trade-off between (1) the physical realizability
and (2) the communication characteristics for a network topology. A
locally and regularly interconnected topology is in general easy to
build and only involves minimal wire and area cost, but it offers poor
global communication characteristics and scales-up poorly with system
size. On the other hand, a random Erd\"os-R\'enyi topology scales-up
well and has a very short-average path length, but it is not
physically plausible because it involves costly long-distance
connections established independently of the Euclidean distance
between the nodes.

\subsection{Addressing Interconnect Challenges by Networks-on-Chip}
The topic of interconnect networks for computers and chips is vast and
complex. Here, we'll give a brief---and certainly
incomplete---overview on communications on chips. From a bird's eye
view, the main challenge of interconnect fabrics consists in
transferring data between two points of the chip with a minimal
latency, minimal energy consumption, and maximal reliability. This job
can obviously be done in a wide variety of ways.  Compared to
computer-to-computer networks, one has to generally deal with a more
restrictive set of resources and with more constraints to
consider. The balance between communication and computation is guided
by numerous design trade-offs and is key to performance. For example,
a set of fast processors is useless if you cannot get enough data to
them on time. 

Traditional VLSI (Very Large-Scale Integration) design uses an ad-hoc
and monolithic communication fabric that connects different resources
(such as for example the memory and the ALU) on the chip together by
dedicated wires. With increasing system complexity and the continuing
miniaturization of the technology, radically new interconnect
approaches will be necessary if we want to sustain the current pace of
progress \cite{meindl03}. Two main factors potentially limit
performance \cite{ho01,davis01}: (1) the miniaturization of wires,
unlike transistors, does not enhance their performance, which is why
wires are now more important than transistors \cite{meindl03}, and (2)
global wires that communicate signals across the whole chip increase
delays and therefore limit the system scalability. The 2005 ITRS
roadmap \cite{itrs2005} (and the 2006 update) lists a more detailed
number of critical challenges for interconnects. In recent years, true
3D architectures and associated design methodologies have emerged,
which offer an attractive option to address some of the current
interconnect challenges \cite{xie06,ahn06}.

On the other hand, networks-on-chip (NoC) \cite{demicheli06,benini02}
have been proposed as a promising solution to address the on-chip
communication challenges and to cope with the increasing communication
requirements. The basic idea behind this paradigm is that the
different modules on the chip (e.g., IP cores) are interlinked by a
bus-like communication network with programmable switch blocks that
support packet-oriented traffic. Thus, NoC architectures decouple the
communication fabric from the processing and storage elements
\cite{benini02} and provide a more modular view of the system, which
allows to better master complexity of large-scale systems, to
decompose it into independent sub-systems, and to keep things
flexible. The additional communication fabric obviously results in an
overhead of area and energy dissipation, which the designer has to
consider in addition to all other design trade-offs. The overhead
largely depends on the connectivity, the number of switch blocks,
their complexity, and the number of possible repeaters. Pande et
al. provide some estimates for their framework \cite{pande05}.

The NoC approach is very general and allows for any interconnect
architecture between functional and communication blocks, such as for
example local and regular, fat-tree, hypercube, or irregular and
application specific, or small-world topologies as described in
Section \ref{sec:complexnets} (see \cite{pande05,benini02} for some
examples). Field-Programmable Gate Arrays (FPGAs), for example, offer
a regular arrangement of programmable logic blocks that are
interconnected by a programmable communication fabric, which
introduces a great degree of freedom for the application designer.
However, once the NoC topology is selected, the only remaining degree
of freedom is the routing strategy.

In the following, we will make use of the NoC paradigm in combination
with nature-inspired 3D network topologies for our explorative
framework. Very recently, other researchers investigated the idea of
inserting long range connections to otherwise regular
networks-on-chip. Ogras et al. \cite{ogras06} showed that a
significant reduction in the average packet latency can be achieved by
superposing a few long-range links to a standard mesh network. In
their approach, the links are however not inserted at random but where
they are most useful. Oshida and Ihara \cite{oshida06} investigate the
performance of a scale-free network-on-chip topologies. Their findings
show that short latencies and a low packet loss ratio can be
achieved. The parallel processing community has also looked into
improving large-scale multi-computer interconnects by adding shortcuts
to bypass nodes \cite{loucif99,dally91}. Fuk\'s and Lawniczak
\cite{fuks99} and Lawniczak et al. \cite{lawniczak03} examined more
generally how the introduction of additional random links influences
the performance of computer networks. For all these approaches, the
performance strongly depends on the routing algorithm and whether it
is able to efficiently use the provided ``shortcuts'' in the networks
while avoiding congestion at the same time.

\section{Description of the Framework}
\label{sec:framework}
We are interested in experimentally exploring self-assembled
networks-on-chip architectures that are built in a largely random
manner. If we want nano-scale electronics to become a success, we have
to show that we can (1) build systems that involve an Avogadro number
of components, and (2) that such a system can efficiently and robustly
solve a specific task. In the absence of mathematical models for
self-assembled electronics (such as for example nanowire growth
models), we decided to build a toy framework that would allow us to
experimentally explore the properties and design trade-offs we are
interested in. The framework also allows us to quantitatively compare
the irregular and self-assembled with representative and regular
nearest-neighbor fabrics.

In the following sections, the network-on-chip-like framework and the
evaluation methodology, which is inspired by Pande et
al. \cite{pande05}, shall be described in more details.

\subsection{Node and Link Model}
\label{sec:nodeslinks}
The basic system-on-chip-like architecture that we use is composed of
(1) programmable computing elements, called {\em processing nodes}
(PNs), and (2) of an associated switch-based interconnect fabric,
which is itself composed of (3) {\em switch nodes} (SNs) and (4)
bi-directional point-to-point interconnects among them. Both
processing and switch nodes can be considered as simple modules of a
large-scale system that need to communicate efficiently among each
other. Figure \ref{fig:topo} shows there different arrangements. The
interconnect topologies shall be described in the next section.

Each switch node can only transmit in parallel messages on $C$
different virtual channels to its neighbors (see e.g. \cite{pande05}
for more details about the concept of virtual channels) according to a
specific routing scheme. No further information processing is done in
the switch nodes.  We assume that they can temporarily store a limited
number of $M$ messages. For the sake of simplicity, we have chosen
this number to be large enough, i.e., $M=100$, to handle our
simulations without creating jamming and losing messages. The
processing nodes, on the other hand, simply send and receive messages
according to a specific traffic scheme. Since we are interested in
interconnect issues here, we do not further specify or limit the
processing nodes' computing capacity.

\subsection{Network Topologies}
\label{sec:topo}
\begin{figure}
  \centering \includegraphics[width=0.45\textwidth]{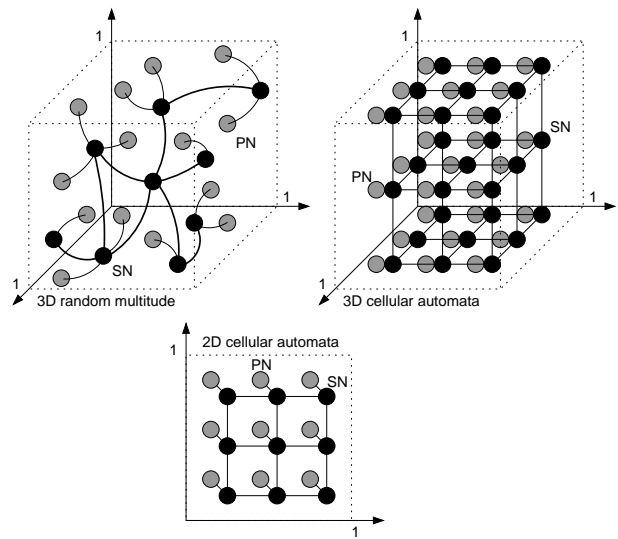}
  \caption{Top left: a 3D random multitude (RM) example composed of
    processing nodes (PNs), switch nodes (SNs), and
    interconnections. Top right: a 3D CA grid-like
    architecture. Bottom: a 2D CA grid-like architecture.}
  \label{fig:topo}
\end{figure}

We have decided to compare the following six reference network
topologies in order to quantitatively evaluate the self-assembled and
irregular fabrics:

\begin{itemize}
  \item {\em 2DCA}: 2D (unfolded) regularly arranged and locally
    interconnected (von Neumann neighborhood), see Figure
    \ref{fig:topo};
  \item {\em 3DCA}: 3D (unfolded) regularly arranged and locally
    interconnected (6 neighbors per switch node), see Figure
    \ref{fig:topo};
  \item {\em 3DRMStandard}: 3D random arrangement, small-world,
    power-law, $\alpha = 1.8$;
  \item {\em 3DRMLocal}: 3D random arrangement, small-world,
    power-law, $\alpha = 3$ (locally interconnected);
  \item {\em 3DRMGlobal}: 3D random arrangement, small-world,
    power-law, $\alpha = 0$ (globally interconnected); and
  \item {\em 3DRMRealistic}: 3D random arrangement, small-world,
    power-law, $\alpha = 1.8$, upper limit $k_{max}$ on the number of
    connections per node, independently of the average connectivity.
\end{itemize}

We call a 3D random arrangement a {\em random multitude} (RM). Figure
\ref{fig:topo} depicts a 3D random multitude, a 2D CA, and a 3D CA
topology. We do not use folded versions for the cellular-automata-like
topologies because that would require long-distance connections. For
both 2D and 3D CAs, the processing nodes are regularly arranged in the
2D or 3D Euclidean space inside a unitary square, respectively
cube. The number of processing nodes is equal to the number of switch
modes, and each processing node is connected to its associated switch
node by a single connection of $0.01$ unit length.

For the random multitude, both processing and switch nodes are
randomly arranged in 3D space, as illustrated in Figure \ref{fig:topo}
(top left). Both the arrangement and the wire topology are inspired by
self-assembled nano-scale electronics, as we will see in Section
\ref{sec:physics}. To make comparisons with the CA grid-like
architectures easier, we assume that each processing node is connected
to the nearest switch node by a single connection only.  The switch
nodes are connected among themselves by a small-world power-law
network \cite{petermann05,petermann06,jespersen00} with average
connectivity $<k_S>$, i.e., each switch node establishes $<k_S>$
connections on average with its neighbors proportional to
$l^{-\alpha}$, where $l$ is the Euclidean distance between the two
switch nodes in question. Thus, the bigger $\alpha$, the more local
the connectivity. For $\alpha = 0$, we obtain the original
Watts-Strogatz small-world topology.  The choice of $\alpha=1.8$ was
guided by the experiments and will be further explained in Section
\ref{sec:alpha}. For all our experiments, we also make sure that the
graphs do not contain disconnected parts.

Algorithm 1 summarizes the construction of a random multitude with
average connectivity $<k_S>$ from an algorithmic point of view. For the
restricted version ({\em 3DRMRealistic}), there is an upper limit
$k_{max}$ on the maximum number of connections per node that can be
established. The idea being this realistic restriction is to make the
topology more physically plausible. More details shall be given in
Section \ref{sec:physics}.

\begin{algorithm}[htb]
\begin{small}
  \label{alg:rmconstr}
  \caption{Construction of a 3D random multitude (RM)}
    \begin{algorithmic}[1]
      \STATE Randomly position $N$ processing nodes within a $1
        \times 1 \times 1$ unit cube (at distinct locations).
      \STATE Randomly position $S$ switch nodes within a $1
        \times 1 \times 1$ unit cube (at distinct locations).
      \STATE $nbLinks$ = $<k_S> \times S$
      \FOR{each processing node $p = 1$ to $N$}
        \STATE Connect $p$ to its nearest switch node.
      \ENDFOR
      \FOR{each link $l = 1$ to $nbLinks$}
        \STATE Randomly choose a switch node $s$.
        \STATE Choose a neighboring node $d$  proportional
        to $l^{-\alpha}$, where $l$ is the Euclidean distance between
        $s$ and $d$
        \IF{``restricted'' multitude type ({\em 3DRMRealistic})}
          \IF{$k_{max}$ reached for node $s$ or $d$}
            \STATE Do not establish a link.
          \ELSE
            \STATE Establish a bi-directional connection between $s$ and $d$.
          \ENDIF
        \ELSE
          \STATE Establish a bi-directional connection between $s$ and $d$.
        \ENDIF
      \ENDFOR
    \end{algorithmic}
\end{small}
\end{algorithm}

\subsection{Routing and Traffic Models}
\label{sec:routing}
Once a topology is chosen and fixed, the only free parameter available
is essentially the routing strategy, which plays a crucial role for
the overall system performance. There exists a large number of
different routing strategies and flavors, which allow to send packets
along pre-specified or dynamically chosen paths in a given network
from a source to a destination. Whether throughput or latency or any
other property needs to be maximized, highly depends on the
application. In general, an ideal routing algorithm is adaptive,
decentralized, robust, and guarantees quality-of-service (QoS) within
well defined bounds.

Here, we deal with packet routing only and the switch nodes do
therefore have to know where to route a packet that they receive. The
path information can be obtained dynamically or statically, based on
the available information on the network's topology. Shortest path
routing is frequently used, but it's not necessary the best routing
strategy \cite{yan06} since congestion has to be taken into account.

Efficient search and information transfer on complex networks while
avoiding congestion and optimizing throughput and latency at the same
time are of great importance in real-world systems
\cite{carmi06,danila06_opt,danila06,lawniczak06,singh03,zhao05,arrowsmith05}. It
has also been shown that small-world and scale-free networks offer
great communication characteristics, are efficient to navigate
\cite{kleinberg00,wang06}, and reduce congestion
\cite{toroczkai04,wu06}. Routing based on local information only
(e.g., \cite{danila06,wang06}) is of particular interest for
large-scale systems where global path information is either not
available or too costly to store in each node.

Here, we are more interested in exploring the extrema than to use any
complicated and highly optimized routing algorithm. We use two main
routing strategies: (1) shortest path routing and (2) random
wandering. Shortest path routing is optimal if the traffic is low and
no congestion occurs, but every node needs to store a routing table
that can get considerably large for large networks. In random
wandering, the switch node that receives a message simply sends it to
a randomly chosen neighbor. This is very simple to implement and
robust against link and node failures, but very inefficient for larger
system sizes. We have also explored ant routing \cite{caro98} as an
alternative, which essentially allows to build shortest paths in a
decentralized manner by the messages (i.e., the ``ants'') themselves.

We decided to adopt a very simple---and admittedly not very
realistic---traffic model, that is however widely used to evaluate
networks: uniform random traffic. Every processing node $n$ injects a
message to a randomly chosen processing node into the network with
probability $p_I$ at each time step. If this injection rate $p_I$ is
$1$, a message will be generated during every time step by every node.

\subsection{Performance Metrics}
\label{sec:perfmetrics}
To compare the different network-on-chip topologies, we use standard
performance metrics and an evaluation methodology inspired by Pande et
al. \cite{pande05}. While area overhead and especially energy
consumption are increasingly important, we are not focusing on these
aspects here. We are mainly concerned by (1) the average number of
hops a message has to take in the shortest path from a source to a
destination, i.e., the number of switch nodes in the path, (2) the
latency, and (3) the shortest path length (in distance
units). Although throughput is important, we assume that for our
applications, we have low traffic that does not lead to congestion.

The average number of hops is measured over $T$ simulation time steps
and over all messages sent. The average shortest path length is
measured in distance units and takes into account the paths between
all possible combinations of processing nodes. The throughput of a
switch node is measured in messages per number of updates per switch
node.

Unless otherwise stated, our experiments used $S=N=64$, a synchronous
node update, $6$ virtual channels per node (i.e., a 3D CA-node could
send a message into all directions simultaneously), an average switch
node connectivity of $<k_S> = 6$, an exact processing node connectivity
of $1$, a traffic injection rate of $p_I=0.1$ (i.e., each node sends a
message every $10^{\textrm{th}}$ time step on average), and a maximum
connectivity of $k_{max} = 10$ for the realistic multitude {\em
  3DRMRealistic}. In our framework, a message is an abstract entity
and we do not take into account its size in terms of number of bits.

\subsection{Physical Plausibility and Realizability}
\label{sec:physics}
There exists an abundance of abstract computing models that are either
hard or impossible (e.g., when infinite resources or time is involved)
to physically realize. Building computers is about hijacking the
underlying material in order to make it do the things we want. Today,
the vast majority of fabrication processes is top-down oriented and
given a computing architecture, the goal is to successfully realize
it, for example by using silicon-based electronics. However, with the
ongoing miniaturization and the constant need for more computing
power, there has been an increasing interest in bottom-up assembling
techniques and disruptive new computing concepts and methodologies.
Especially self-assembling molecular electronics, based for example on
nanowires or nanotubes, bears unique challenges and opportunities for
new paradigms.

Despite important progress, the fabrication of ordered 3D hierarchical
nano-structures remains very challenging \cite{zhang06}. Here, we
argue that because of fewer physical constraints, computing
architectures that are ``assembled'' in a largely random manner, are
easier and cheaper to build than highly regular architectures, such as
crossbars or cellular-automata-like assemblies, which usually require
a perfect or almost perfect establishment of the
connections. Self-assembly, for example, is particularly well suited
for building random structures \cite{zhirnov01}. Power-law
connection-length distributions have been observed in many systems
created through self-organization, such as the human cortex or the
Internet, and they can be considered ``physically realizable''
\cite{petermann06}. Such topologies evolve naturally in Nature,
essentially because of the cost associated with long distance
connections, which prevents a uniform wiring probability over all
nodes. 

There is very little work about computing architectures with irregular
assemblies of connections and components. Tour et al.  \cite{tour02},
for example, explored the possibility of computing with randomly
assembled, easily realizable molecular switches, that are only locally
interconnected, however. On the other hand, Hogg et al. \cite{hogg06}
present an approach to build reliable circuits by self-assembly with
some random variation in the connection location.  At the exception of
a few researchers (e.g., \cite{patwardhan06,lawson06,tour02}), the
vast majority working in the field of nano-scale electronics tries to
build regular structures, which allow for a more or less
straightforward mapping of higher-level computing
architectures. Computing with highly irregularly assembled physical
substrates is undoubtedly a new and disruptive paradigm. The main
question we would like to address to some extent in this section is
whether and how the framework as described above could be physically
implemented.

Designing nanoscale interconnects is guided by a number of major and
dependent trade-offs: (1) the number of long(er)-distance connections,
(2) the physical plausibility, and (3) the efficiency of
communication. Since the fabrication of nano-scale computing
architectures tends to be very challenging, we opt for a
fabrication-friendly approach and will try to live with what we can
currently build.  Although we are unable to provide experimental
results at this point, plausible approaches for physical realizations
shall be briefly sketched here. Preliminary experiments with both
nanowires and nanotubes are currently underway at the Los Alamos
National Laboratory.

We believe that a random multitude would be best realized in a hybrid
way today, where the processing and switch nodes are for example made
of current (nanoscale) silicon. Since we are focusing on the
interconnections here, we do not further specify the characteristics
of these nodes. Our only intention is to keep both the computation and
the routing as simple as possible to minimize the node's complexity.
The interconnect fabric would then be gradually self-assembled using
either nanowires or nanotubes \cite{hu99}. We imagine that in a first
step, both the processing and the switch nodes would either be
randomly placed in a scaffolding structure or be submerged in some
kind of fluid, similar to the self-assembly of nanowires from a
solution \cite{cheng05}. Each processing and switch node will have a
limited number of seed points, where either nanowires or nanotubes
could connect to or grow out by, for example, a self-catalytic growth
process. The wires would grow in random directions and eventually make
contact with another switch or processing node. The growth could be
guided by additional scaffolding structures or for example by
magnetical and electrical fields. Ye et al. \cite{ye06}, for example,
present an approach for the directed assembly by means of
electrodeposition or vapor deposition. As an alternative to directly
growing wires from seed points on the nodes, one could also imagine to
pre-fabricate the wires and then to connect them in a second step to
the nodes, for example while being immersed in a solution.

To obtain a specific power-law distribution of connection lengths, and
thus to obtain a small-world topology as described above, one could
imagine to grow different wires with different probabilities, whose
lengths follow a power-law distribution. Alternatively, if the wires
directly grow out of the nodes and randomly connect to neighboring
nodes, we hypothesize that it is possible to obtain a desired
distribution as a function of the Euclidean distance between the nodes
by imposing restrictions on the wire growth lengths. However, physical
wire growth models or experimental results would be necessary to
further investigate this option.  Note also that current nanowires
tend to be rather short because of a high resistance and the
probability of breaks, which will naturally limit the number of
long-distance connections.

In order to make our framework as realistic as possible, we introduced
in Section \ref{sec:topo} a limitation $k_{max}$ on the number of
links that a node can carry for the {\em 3DRMRealistic} random
multitude. Given the above growth mechanisms, this is a realistic
assumption because one cannot assume an unlimited number of contacts on
a given node. The exact value of $k_{max}$ will depend on the
wire-type and the fabrication technology. For all our experiments, we
use a value of $k_{max} = 10$, which seems rather pessimistic, but
allowed to make a plausible comparison with the unrestricted random
multitude, {\em 3DRMStandard}.

In summary, we believe that there exist several promising paths to
physically realize irregularly self-assembled networks of wires and
nodes with a specific topology. The random multitude construction
algorithm (see Section \ref{sec:topo} of our framework is meant to
reflect what we could possible assemble in reality in the very near
future.

\section{Experiments}
\label{sec:experiments}
In the following sections, we'll perform a number of pragmatic and
simple experiments with the goal to compare the performance of
realistic, both regular and irregular network-on-chip topologies as
presented in Section \ref{sec:topo}. All simulations were written in
Matlab. 

\subsection{Experiment 1: System Scalability}
The goal of this first experiment is to illustrate how the different
topologies perform as the system size scales up. What works for $N=64$
nodes does not necessary work for $N=10000$ nodes. As we have seen
before, scalability is a critical issue for nano-scale systems because
it is generally very easy to build systems that involve huge numbers
of components, e.g., Avogadro-scale systems. If the communication
fabric doesn't allow to efficiently send data across such an assembly,
it will be impossible to solve tasks efficiently and thus to stay
competitive with conventional design approaches.

For all six assemblies as described in Section \ref{sec:topo}, we have
varied the system size and measured the average number of hops, which
is proportional to the average path length $L$, i.e., the number of
edges in the shortest path between two nodes, averaged over all pairs
of nodes, as for example used in \cite{watts98}. The different system
sizes we used were: (1) $N=S=[9,14,19,24,29,34,39,44,49,54,59,64]$ for
random multitudes, (2) $N=S=[9,16,25,36,49,64,81,100,121]$ for 2D CAs,
and (3) $N=S=[8,27,64,125]$ for 3D CAs. Shortest path routing was
used.

\begin{figure}
  \centering \includegraphics[width=0.45\textwidth]{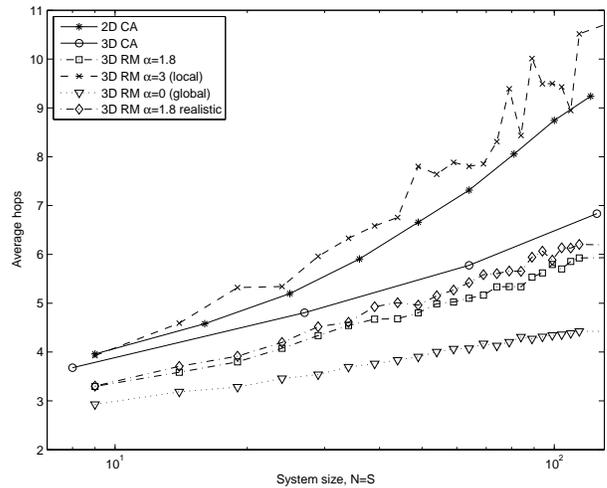}
  \caption{Scaling of the number of hops as a function of the system
    size $N=S$. The $\alpha=0$ (global), $\alpha=1.8$ (standard), and
    $\alpha=1.8$ (realistic) random multitudes show a logarithmic
    scaling behavior. Average over $10$ runs for RMs and over $7$ runs
    for CAs.}
  \label{fig:hopslog_bw}
\end{figure}

As Figure \ref{fig:hopslog_bw} shows, the locally connected
topologies, i.e., the 2D and 3D CA as well as the local 3D random
multitude, scale up worse with system size than the other three
topologies. Not surprisingly, the globally connected random multitude
($\alpha = 0$) scales up best because of the uniform rewiring
probability over all nodes, independently of the Euclidean distance
between them. The average path lengths of small-world graphs scale up
logarithmically with the number of nodes, which Figure
\ref{fig:hopslog_bw} confirms. Note that there is only a very small
difference between the realistic random multitude and the unrestricted
$\alpha = 1.8$ multitude. This is good news and illustrates that the
limited and thus more realistic connectivity has little effect on the
average number of hops as the system is scaled up.

\subsection{Experiment 2: Local versus Global Connections}
\label{sec:alpha}
The distribution of the long- and short-distance connections as a
function of the Euclidean distance between the nodes is a crucial
parameter in our model. Clearly, we are interested in a great network
performance while having a minimal number of global connections, which
are generally costly to establish.  In this experiment, we explore the
network's communication characteristic as a function of the parameter
$\alpha$ and compare it with the fixed 2D and 3D CA grid-like
assemblies. The function $l^{-\alpha}$, where $l$ is the Euclidean
distance between the two switch nodes in question, describes the
connectivity of the random multitudes.

\begin{figure}
  \centering \includegraphics[width=0.45\textwidth]{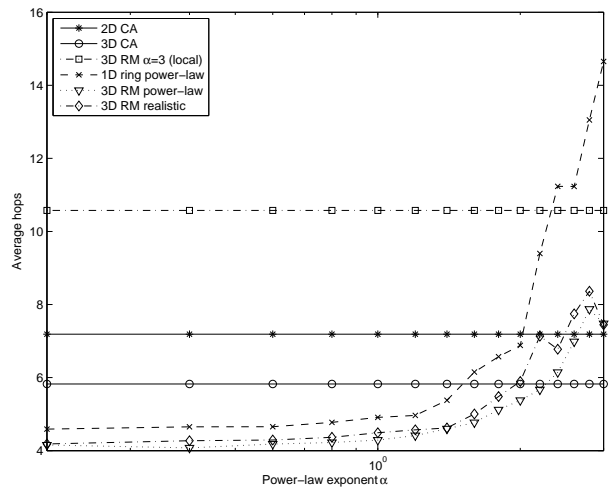}
  \caption{The average number of hops as a function of the power-law
    exponent $\alpha$. Average over $2$ runs, shortest path routing,
    $N=S=64$.}
  \label{fig:alphahopslog_bw}
\end{figure}

Figure \ref{fig:alphahopslog_bw} shows the average number of hops as a
function of the power-law exponent $\alpha$. The 2D and 3D CA as well
as the locally connected and fixed random multitude ($\alpha=3$) are
plotted as a reference, although their value is independent of
$\alpha$. As one can see, the number of hops increases dramatically
the bigger $\alpha$ gets, i.e., the more local the connectivity
becomes. For comparison, the 1D ring structure as used in the original
Watts-Strogatz rewiring model \cite{watts98} is also shown. The 1D
ring structure performs worse with increasing $\alpha$ than the 3D
random multitudes, which offer a higher connectivity. For a value of
$\alpha$ that is slightly smaller than $2$, both the unrestricted and
the realistic random multitude perform better than the 3D regular
assembly.

\begin{figure}
  \centering \includegraphics[width=0.45\textwidth]{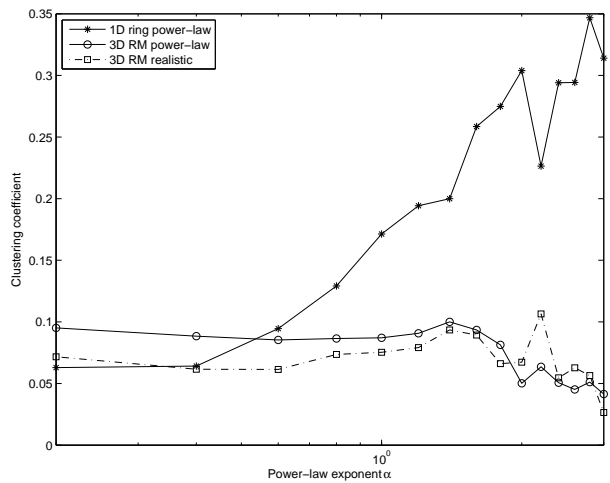}
  \caption{The clustering coefficient as function of the power-law
    exponent $\alpha$. Average over $2$ runs, shortest path routing,
    $N=S=64$.}
  \label{fig:alphaclusterlog_bw}
\end{figure}

In our framework, the average latency is essentially proportional to
the average number of hops because we keep the traffic injection rate
low to avoid jamming. Figure \ref{fig:alphahopslog_bw} also confirms
the small-world characteristic of the wiring, where the average path
length---the average number of hops in our case---drastically drops
when a few global connections are added (i.e., when $\alpha$ becomes
smaller).  Figure \ref{fig:alphaclusterlog_bw} shows the clustering
coefficient $C$ \cite{watts98} of the 1D ring and the two 3D random
multitudes. The random multitudes have very low clustering
coefficient, while the 1D ring behaves like the Watts-Strogatz model.

As Petermann and De Los Rios \cite{petermann06} find both analytically
and numerically, the small-world phenomena in a network built using a
power-law decaying distribution of shortcut lengths occurs when $\alpha
< D + 1$, where $D$ is the network's dimension. In the case of our
random multitudes, $D=3$, which confirms our observations of
small-world behavior. Based on the results as shown in Figure
\ref{fig:alphahopslog_bw}, we have chosen $\alpha=1.8$ for the regular
random multitudes ({\em 3DRMStandard} and {\em 3DRMRealistic}) for the
following experiments.

\subsection{Experiment 3: Number of Switch Nodes and  Connectivity}
From a design perspective, one is obviously interested to minimize the
number of switch nodes and the connectivity among the switch nodes to
a level that doesn't significantly affect the system performance. In
this experiment, we explore the framework's characteristics by varying
the number of switch nodes $S$ and the switch node connectivity
$<k_S>$, while keeping $\alpha$ constant.

\begin{figure}
  \centering \includegraphics[width=0.45\textwidth]{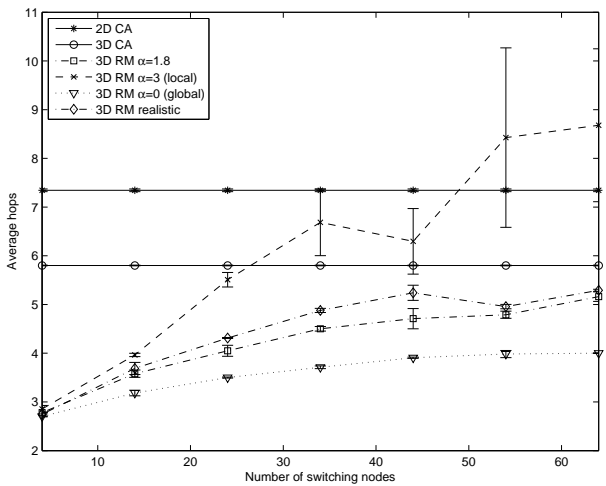}
  \caption{The average number of hops as a function of the number of
    switch nodes $S$. Average over $2$ runs, shortest path routing, $N=64$.}
  \label{fig:shops_bw}
\end{figure}

Figure \ref{fig:shops_bw} shows that the average number of hops for a
message to take on the shortest path from any source to any
destination increases with the number of switch nodes $S$. The fixed
CA topologies are shown for comparison. However, there is a trade-off
between the number of hops and the throughput a network can handle. A
low number of switch nodes limits the throughput, which we cannot
illustrate here because we have chosen a low traffic injection rate
that avoids jamming.  Thus, depending on what the network needs to be
optimized for (i.e., lower number of hops, throughput, short average
path length, etc.), one can make the appropriate choice for the number
of switch nodes. Obviously, the amount of hardware resources and the
volume required will also come into play in reality.

\begin{figure}
  \centering \includegraphics[width=0.45\textwidth]{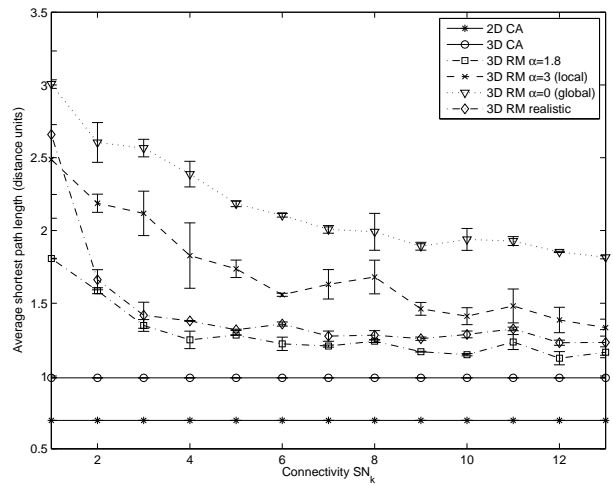}
  \caption{The average path length as a function of the average switch
    node connectivity $<k_S>$. Average over $2$ runs, shortest path
    routing, $N=S=64$.}
  \label{fig:snkpath_bw}
\end{figure}

Figure \ref{fig:snkpath_bw} shows that the higher the switch node
connectivity $<k_S>$, the lower the average shortest path length. The
fixed CA topologies are shown for comparison. Once again, one can
observe only a small difference between the unrestricted and the
realistic random multitude. Further results shall be summarized here:

\begin{itemize}
  \item A higher switch node connectivity decreases both the average
    latency and the average number of hops. The throughput is only
    slightly improved.

  \item The higher the number of switch nodes $S$, the higher the
    number of hops and the higher the average latency. The lower $S$,
    the higher the average path length and the higher the throughput.

  \item The higher the number of virtual channels $C$, the higher the
    node throughput (within the limits of the capacity of the physical
    links) and the lower the average latency. The average shortest
    path length is not affected by $C$.
\end{itemize}

We can conclude that there are no ``optimal'' values for connectivity,
the number of switch nodes, and the number of virtual
channels. Instead, choosing the right values is a matter of dependent
trade-offs in the design space. Local connections are very interesting
from an implementational point of view, but offer reduced global
communication characteristics only, which directly affects the efficiency
of problem solving (see also Section \ref{sec:task}). Adding a few
long(er)-distance connections proportional to the distance between the
nodes is physically plausible and greatly improves the overall system
performance as well as the robustness, as we will see in the next
section.

\subsection{Experiment 4: Robustness against Link Failures}
Robustness against manufacturing defects and dynamic failures is
critical for future Avogadro-scale self-assembled nano-scale
architectures \cite{dehon05}. Due to the small feature sizes, such
systems are generally expected to be much more prone to
radiation-induced soft errors, to thermal noise \cite{kish02}, and to
fabrication defects because of the self-assembly processes.

\begin{figure}
  \centering \includegraphics[width=0.45\textwidth]{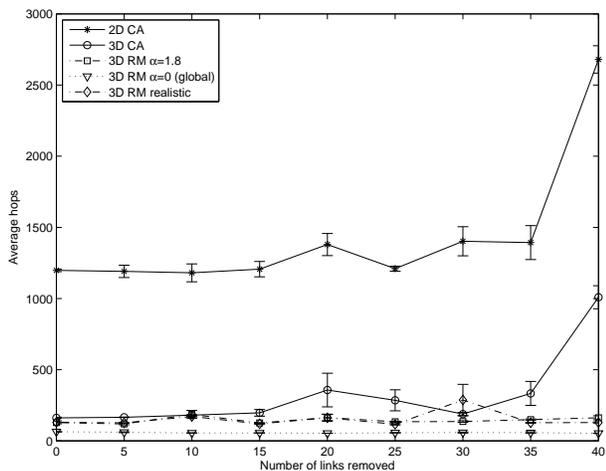}
  \caption{The average number of hops as a function of the number of
    randomly deleted links. Average over $2$ runs, random wandering,
    $N=S=64$.}
  \label{fig:del_bw}
\end{figure}

In order to compare the robustness against link failures of both
regularly and irregularly interconnected topologies, we randomly
removed links between the switch nodes in our six reference
assemblies. It is well known that small-world and scale-free networks
are robust against random failures of nodes and links
\cite{albert02,petermann05}. As Figure \ref{fig:del_bw} illustrates,
this is also valid for our framework. The average number of hops is
plotted as a function of the number of removed switch links. In this
experiment, we used random wandering to illustrate an extreme case. As
one can see, both 2D and 3D CA-topologies start to perform worse,
i.e., the average number of hops increases, when the number of
randomly removed links approaches $40$, while the random multitudes
essentially remain unaffected by the removed links.

The random link failures admittedly represent a well oversimplified
fault model, nevertheless, it illustrates that the irregular
small-world topologies of our framework provide ``robustness for
free'' to some extent, even without using any specific fault detection
and isolation technique.

\subsection{Experiment 5: Solving a Simple Task}
\label{sec:task}
In this last experiment, we are interested in evaluating the
performance of solving a simple problem, which is well-known in the
area of cellular automata (CA): the {\em synchronization task}. This
``global'' task is essentially trivial to solve if one has a global
view on the entire system (i.e., if one has access to the state of all
nodes at the same instant in time), but it is non-trivial to solve for
locally connected cellular automata or random boolean networks
(RBNs). Although it is commonly called a ``toy problem,'' the
synchronization task has actually various real-world applications,
such as for example in sensor networks, where one cannot assume global
synchronization and global signals, and thus special mechanisms are
required \cite{li06}.

The {\em synchronization task} (also called {\em firefly task}) for
synchronous CAs was introduced by Das et al.  \cite{dasetal95} and
studied among others by Hordijk \cite{hordijk96} and Sipper
\cite{sipper97:book}. In this task, the two-state $D$-dimensional
automaton, given any initial configuration, must reach a final
configuration within $M$ time steps, that oscillates between all $0$s
and all $1$s on successive time steps. The whole automaton is then
globally synchronized.

Here, we use a slightly adapted version the task for our framework: we
assume that each processing node in our framework contains an
oscillator which frequency is specified by a number between $0 \leq
f_{osc} \leq 1$. The modified task then consists to find a common
frequency for all oscillators. The algorithm is inspired by the {\em
  averaging algorithm} as described in \cite{li06}. Each processing
node state is initialize to a random value from the interval $[0$,$1]$
before it repeatedly performs the following steps in an asynchronous
manner: (1) send current oscillator frequency to a random processing
node; (2) if the current node $i$ receives a message from any other
processing node $r$, then average own oscillator $f_i$ with neighbor
frequency $f_r$; (3) set own oscillator to this frequency $f_i =
\frac{f_i + f_r}{2}$; and (4) also send it to a new random processing
node.

\begin{figure}
  \centering \includegraphics[width=0.45\textwidth]{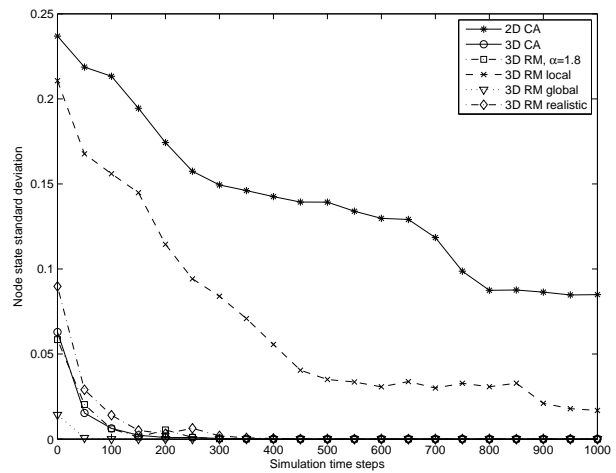}
  \caption{Performance of the synchronization task. The smaller the
    standard deviation of the node state values, the better the nodes
    are synchronized. The initial values for each curve depend on the
    randomly initialized network state. $N=S=64$, random wandering.}
  \label{fig:syncstd_bw}
\end{figure}

There are obviously numerous (also more efficient) ways to solve this
task, but here we are interested in an illustrative comparison rather
than in the absolute performance values and limits. We compared how
this simple algorithm performed on the investigated interconnect
fabrics by using random wandering. As Figure \ref{fig:syncstd_bw}
illustrates, the small-world topologies perform best. Both the 2D and
the locally interconnected multitude very slowly converge because of
the poor global communication characteristics. Not surprisingly, the
globally connected random multitude performs best.

It has been shown elsewhere that irregular small-world interconnects
perform better on both the synchronization (e.g.,
\cite{guclu06,mesot05,hong04,motter05,motter05}, and many others) and
the density classification task (e.g., \cite{mesot05}) than purely
locally interconnected topologies. However, the frameworks and
assumptions used in each approach are somehow different and sometimes
not straightforward to compare. The results of our framework merely
confirm what has been found theoretically elsewhere and in our two
previous experiments, namely that the excellent communication
characteristics (i.e., short characteristic path length, the small
latency, etc.) also helps to efficiently solve tasks, especially tasks
which require a lot of global communication. From an evolutionary
perspective, this also seems the reason why most natural networks,
e.g. the brain \cite{hellwig00,egueluz05,sporns04}, have evolved with
the small-world and scale-free property. The relationship between
efficient problem solving and interconnection topologies has naturally
also preoccupied the parallel computing community since its beginning
(e.g., \cite{levitan87}).

\section{Conclusion}
\label{sec:conclusion}
We have experimentally investigated in a pragmatic way several
relevant metrics of both regular and irregular, realistic
system-on-chip-like computing architectures for self-assembled
nanoscale electronics, namely 2D and 3D local-neighborhood as well as
irregularly build small-world interconnects with different
distributions of long-distance connections. The small-world
architectures with a power-law decaying distribution of shortcut
lengths investigated in our framework are both physically plausible,
could likely be built very economically by self-assembling processes,
possess great communication characteristics, and are robust against link
failures. While regular and local-neighborhood interconnects are
easier and more economical to build than interconnects with lots of
global or semi-global long-distance connections, we have seen in the
previous section that they are not as efficient for global
communication, which is very important and directly affects how
efficient problems can be solved with such architectures. Small-world
networks with a uniform distribution of long-distance connections or
pure Erd\"os-R\'enyi random networks, on the other hand, are not
physically plausible because one has to assume an increasing wiring
cost with distance. As our results have shown by means of a
simplistic, yet realistic framework, small-world topologies with a
power-law decaying distribution of shortcut lengths offer a unique
balance between performance, robustness, physical plausibility, and
fabrication friendliness. In addition, it has been shown that adaptive
routing---which we have not explored in detail here---is very
efficient on small-world power-law graphs \cite{kleinberg00,wang06}.

We believe that computation in random self-assemblies of components
and interconnections (see e.g.,
\cite{tour02,lawson06,hogg06,patwardhan06}) is a highly appealing
paradigm, both from the perspective of fabrication as well as
performance and robustness. This is obviously a radically new
technological and conceptual approach with many open questions. For
example, there are essentially no methodologies and tools that would
allow (1) to map an arbitrary computing architecture or a logical
system on a randomly assembled physical substrate, (2) to do arbitrary
computations with such an assembly, and (3) to systematically analyze
performance and robustness within a rigorous mathematical
framework. There are also many open questions regarding the
self-assembling fabrication techniques, which will need to be further
explored in the future.

\subsection{Acknowledgments}
\label{sec:ack}
The author is very grateful for the fruitful discussions with Elshan
A. Akhadov, Steven K. Doorn, Sandip Niyogi, Tom Picraux, and Evsen
Yanmaz.

\end{document}